\documentclass{article}
\usepackage[T1]{fontenc} % add special characters (e.g., umlaute)
\usepackage[utf8]{inputenc} % set utf-8 as default input encoding
\usepackage{ismir,amsmath,cite,url}

\usepackage{graphicx}
\usepackage{color}
\usepackage{booktabs}
\usepackage{colortbl}
\usepackage{lineno}
\usepackage{soul}
\title{Parameter Sensitivity of Deep-Feature based Evaluation Metrics for Audio Textures}

\multauthor
{Chitralekha Gupta \hspace{1cm} Yize Wei \hspace{1cm} Zequn Gong} { \bfseries{Purnima Kamath \hspace{1cm} Zhuoyao Li \hspace{1cm} Lonce Wyse}\\
National University of Singapore\\
{\tt\small chitralekha@u.nus.edu, yize.wei@u.nus.edu, zequn.gong@u.nus.edu}\\ {\tt\small purnima.kamath@u.nus.edu, zhuoyaoli@u.nus.edu, lonce.wyse@nus.edu.sg}
}
\def\authorname{C. Gupta, Y. Wei, Z. Gong, P. Kamath, Z. Li, and L. Wyse}

\begin{document}

\maketitle
\begin{abstract}
Standard evaluation metrics such as the Inception score and Fréchet Audio Distance provide a general audio quality distance metric between the synthesized audio and reference clean audio. However, the sensitivity of these metrics to variations in the statistical parameters that define an audio texture is not well studied. In this work, we provide a systematic study of the sensitivity of some of the existing audio quality evaluation metrics to parameter variations in audio textures. Furthermore, we also study three more potentially parameter-sensitive metrics for audio texture synthesis, (a) a Gram matrix based distance, (b) an Accumulated Gram metric using a summarized version of the Gram matrices, and (c) a cochlear-model based statistical features metric. These metrics use deep features that summarize the statistics of any given audio texture, thus being inherently sensitive to variations in the statistical parameters that define an audio texture. We study and evaluate the sensitivity of existing standard metrics as well as Gram matrix and cochlear-model based metrics to control-parameter variations in audio textures across a wide range of texture and parameter types, and validate with subjective evaluation. We find that each of the metrics is sensitive to different sets of texture-parameter types. This is the first step towards investigating objective metrics for assessing parameter sensitivity in audio textures. 
\end{abstract}
\section{Introduction}
\label{sec:intro}
Audio textures are rich and varied sounds that are produced by a superposition of multiple acoustic events. Unlike the sound of an individual event, such as a footstep, or the complex spectrotemporal structures and sequences of speech or music, an audio texture is defined by properties or parameters that remain constant over time \cite{mcdermott2011sound} despite statistical variation over time or between instances. 
% \textcolor{Brown}{Therefore, the goal of parametric synthesis algorithms is to synthesise audio textures based on a set of parameters that can describe the texture. If these parameters are chosen well, any two textures with the same parameter values would sound like the same kind of texture without necessarily being identical. For example, two audio snippets of tapping, both at the same rate of tap occurrence, should be considered as the same texture, even if the snippets are not temporally identical, i.e.~the taps do not occur at the same time instance in the two snippets. Hence,} 
For parametric audio texture synthesis, the goal is to generate a sound with descriptive parameters that match a target texture's parameters. Standard objective evaluation metrics for audio texture synthesis include diversity and quality based measures such as the Inception score \cite{engel2017neural}, and Fréchet Audio Distance (FAD) \cite{kilgour2019frechet}. However, the existing metrics have not been studied for their sensitivity to systematic variation in the parameter values that define the synthesized audio texture.

Various audio texture synthesis algorithms have been applied to modeling a wide range of natural audio textures including rhythmic (eg.~drums, tapping), pitched (eg.~windchimes, churchbells) and other natural sounds (eg.~rain, wind), but with different degrees of success. Evaluation of parametric variation textures has typically been through human listening experiments \cite{mcwalter2018adaptive, antognini2019audio}.

% The quality and flexibility of the synthesis entirely depend on the choice of parameters.
``Deep features'' (activation patterns across layers of neural networks) have been explored for various evaluation tasks such as out-of-distribution detection in audio classification \cite{sastry2019detecting}, perceptual image similarity evaluation \cite{zhang2018unreasonable}, audio distortion assessment \cite{kilgour2019frechet} and audio texture synthesis \cite{antognini2019audio, ulyanov2016audio, mcdermott2011sound}. In this work  
% \textcolor{Brown}{ we ask the question, can we use deep features to design a metric that is sensitive to parameter variations in a synthesized audio texture? Such a metric would be useful to evaluate, for example, whether or not the intended tempo of a drum break audio texture is actually preserved in the synthesized audio, or has it become too fast or too slow. It could also be used to evaluate whether the wind strength in the synthesized wind sound deviates from the intended strength. One may argue that there could be a specific beat tracking algorithm, or a wind-strength assessment algorithm to track and evaluate the correctness of these parameters. However, such a strategy would be specific to the sound type and would fail to generalize over a wide range of sounds. In this work, we explore potential evaluation metrics that are generally sensitive to parameter variations and hence, suitable for assessing parameter fidelity in the synthesized audio texture. }
we study several existing objective metrics and also explore deep features and cochlear channel model statistics for potential evaluation metrics sensitive to parameter variations.
% We also explore cochlear channel based texture statistics \cite{mcdermott2011sound} for comparison. We hypothesize that these deep features would enable objective comparisons between audio textures. 
We make use of a controlled audio texture dataset \cite{Wyse2022Syntex} to study these metrics for a wide range of audio textures - rhythmic, pitched, and non-rhythmic non-pitched under systematic parametric variation. We present a comparative study of the parameter sensitivity of the metrics and validate their reliability through human listening tests.

% In Section \ref{sec:relatedwork}, we discuss the previous related work. In Section \ref{sec:metrics}, we describe the existing metrics that we compare against as well as our proposed metrics. In Section \ref{sec:experiments}, we elaborate on our experimental setup including the dataset, architectural description, and human evaluation. And finally, in Section \ref{sec:results}, we present our results and findings, and we conclude in Section \ref{sec:conclusion}.
% \vspace{-0.5cm}
\section{Related Work}
\label{sec:relatedwork}
\subsection{Why is a parameter sensitive metric needed?}
%In this subsection, describe in detail how many synthesis models fail to produce the parameter correctly, hence justifying the reason why a metric for evaluating parameter fidelity is important.
McDermott and Simoncelli~\cite{mcdermott2011sound} developed a set of statistics based on a cochlear model to describe the perceptually relevant aspects of a given audio texture. To synthesize a new audio texture, a random input is iteratively perturbed until its statistics match those of a target. This algorithm produces convincing audio for many natural textures such as insect swarms, a stream, or applause. However, human listeners give low scores on realism for resynthesized pitched and rhythmic textures, such as wind chimes, drum break, walking on gravel, and church bells. Besides the audio quality, the statistical parameters associated with these sounds can fail to be faithfully preserved in the synthesized sounds\cite{mcwalter2018adaptive}. Moments derived from the time-averaged scattering coefficients when used for resynthesis of audio textures have also shown similar performance but using fewer coefficients\cite{bruna2013audio,anden2019joint}. An objective metric for predicting these human evaluations of texture synthesis failures would be valuable.

%%%Besides the audio quality, the statistical parameters associated with these sounds \textcolor{violet}{can fail to be} faithfully preserved in the synthesized sounds. However, the only way to realize that was through human listening test.
%Include mc walter's paper here

Gatys et al.~\cite{gatys2015texture} did seminal work on image texture synthesis that replaced hand-crafted statistics with Gram matrix statistics computed as the correlation between hidden feature activations from layers of a trained convolutional neural network (CNN). Iteratively perturbing a random input to match Gram matrices produces compelling and novel image textures. Similarly, Ulyanov et al.~\cite{ulyanov2016audio} improved the quality of synthetic textures based on the dataset from McDermott and Simoncelli \cite{mcdermott2011sound}. Antognini et al.~\cite{antognini2019audio} extended Ulyanov's work by modifying the architecture with 6 parallel single-layered untrained CNNs each with a different convolutional kernel size, and included autocorrelation and diversity losses in addition to the original gram matrix loss. They demonstrated some improvements, but found failure modes similar to that of McDermott and Simoncelli~\cite{mcdermott2011sound} using a combination of objective and subjective evaluation. Caracalla and Roebel~\cite{caracalla2020sound} also relied on human listening tests to show the improvement in sound quality achieved in this kind of iterative texture synthesis using a complex spectrogram audio representation.  

The key take-away from these previous studies is that there is a need for evaluating the preservation of statistical parameters in synthesised audio texture but a lack of an objective metric for that purpose.

% \textcolor{red}{Is this kind of "header" legal?}
\vspace{-0.2cm}
\subsection{The existing audio synthesis evaluation metrics}

The evaluation of generative models in terms of perceptual realism is challenging. However, various objective metrics have been formulated for assessing the quality of synthesized audio textures. For example, L2 distance, cosine distance, and signal-to-distortion ratio use a clean reference signal to compare with the modified or enhanced synthetic signal. Such signal-level metrics are agnostic to the type of audio that is being enhanced. However, these metrics may not always correlate with human perception. FAD \cite{kilgour2019frechet} takes a different approach as a reference-independent metric that, instead of looking at individual clips, computes the distance between the distribution of embedding statistics generated on a large set of clips.

The Inception score \cite{salimans2016improved,engel2017neural,engel2019gansynth}, passes generated examples through a pre-trained classifier. The mean KL divergence between the conditional output class probabilities and the marginal distribution of the same are then calculated. The Inception score penalizes models whose examples are not easily classified into a single class, as well as models whose examples collectively belong to only a few of the possible classes. However, the Inception score is not the right tool for evaluating a model's sensitivity to parameter differences between sounds within a class.   

The goal of most of these audio quality assessment metrics has been to quantify the degradation of an enhanced/modified audio signal. However, to the best of our knowledge, there is a lack of a systematic study of the sensitivity of these metrics to parameter variations in synthesized audio. Such a study is particularly challenging for audio texture synthesis because of the inherent variations that correspond to a particular parametric description.

\section{Metrics}
\label{sec:metrics}
We study two standard metrics and three Gram matrix metrics for sensitivity to parametric changes to audio textures.
\subsection{Existing Metrics}
\subsubsection{Fréchet Audio Distance (FAD)}
FAD \cite{kilgour2019frechet} is a reference-independent metric for audio quality assessment that computes the Fréchet distance \cite{dowson1982frechet} between the multi-variate Gaussian distributions of the embeddings of train and test set audio data. These 128 dimensional embeddings are extracted from a VGG-ish model \cite{hershey2017cnn} pre-trained on clean audio data for classification. 
\begin{equation}
    FAD = ||\mu_b-\mu_t||^2+tr(\Sigma_b+\Sigma_t-2\sqrt{\Sigma_b\cdot \Sigma_t})
\label{eq:FAD}
\end{equation}
where the training and test data embeddings are assumed to have multivariate Gaussian distributions $\mathcal{N}(\mu_b,\Sigma_b)$ and $\mathcal{N}(\mu_t,\Sigma_t)$, respectively. 
% The VGGish is trained on a large dataset of YouTube videos as an audio classifier with over 3000 classes.
% The activations from the 128 dimensional layer prior to the final classification layer are used as the embedding. 
We used the open-source model and FAD computation code. %\footnote{\url{https://github.com/google-research/google-research/tree/master/frechet_audio_distance}} are open-source. %\footnote{\url{https://github.com/tensorflow/models/tree/master/research/audioset}} 

% In all the experiments in this paper, we generate a set of ten audio clips with the same control parameter values, so that FAD can be computed over the distribution of the embeddings of these ten clips. Although large number of clips are recommended for computing unbiased distribution estimates for FAD calculation \cite{chong2020effectively}, in this work, we chose ten and used the same amount of data per audio texture type and control parameter value combination. Therefore, the trends being observed in our experiments are consistent and comparable across different texture types and parameters, although the absolute values may change if the amount of data is scaled up.

\subsubsection{L2 Distance}
We compute the magnitude L2 distance between spectrograms as another standard metric for our study, as
\begin{equation}
    L2 = |||stft(x_A)| - |stft(x_B)| ||_2
\end{equation}
where, $x_A$ and $x_B$ are the two signals to be compared.
% \section{Gram matrix Metric}
% \subsection{Definition}
\vspace{-0.2cm}
\subsection{Gram matrix based Metrics}
\subsubsection{Gram matrix Metric (GM)}
Following Ulyanov and Lebedev \cite{ulyanov2016audio} and Antognini et al.~\cite{antognini2019audio}, we use 2D spectrogram representations of audio for iterative updates through the CNNs used to compute a summary statistic in the form of a Gram matrix. Two audio textures sound similar if the computed Gram matrices are similar. Here, we leverage on this property of Gram matrix to develop a metric designed to be sensitive to parameter variations. We define the Gram matrix metric as the mean squared error between the Gram matrices of two textures. We hypothesize that this metric would be sensitive to parametric variations in the statistics of the same texture type such as different rates of flowing water.

\begin{figure}
    \centering
    \includegraphics[width=0.75\columnwidth]{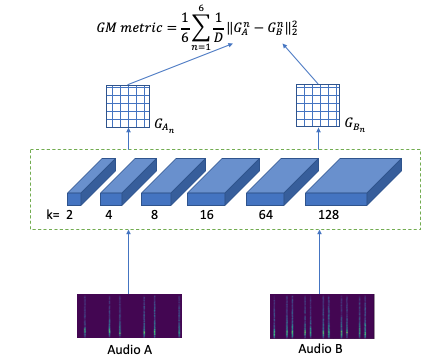}
    \caption{Gram matrix param-sensing metric computation block diagram consisting of an ensemble of six 1 layer 1d-CNNs, with different kernel sizes, k. Each CNN has 512 filters, and the weights are randomly initialized.
}
    \label{fig:grammetric}
\end{figure}

Formally, we define the Gram matrix metric $GM$ as,
\begin{equation}
    GM=\frac{1}{N}\sum_{n=1}^{N}\frac{1}{D}||G_{A}^n-G_{B}^n||_2^2
\end{equation}
where, $G_{A}^n$ and $G_{B}^n$ are flattened Gram matrices derived from audio clips A and B respectively for the $n^{th}$ CNN in an ensemble of $N$ CNNs, and $D$ is the total number of elements in the Gram matrix. In this work, we adopted the architecture used by Antognini et al.~\cite{antognini2019audio} for the Gram matrix computation, as shown in Figure \ref{fig:grammetric}. It consists of an ensemble of 6 ($N$=6) single-layered CNNs. We used spectrograms of the audio texture clips as the input to the network computed with 512 FFT bins and hop-size of 128 samples. Unlike images, we consider the spectrograms as one-dimensional input features with the frequency bins as the number of channels. We therefore use a one-dimensional convolution, where each CNN has a convolutional kernel $k_n$ with a different width, in particular 2, 4, 8, 16, 64, 128. The different kernel sizes capture statistics over multiple time scales. We used $F=512$ filters randomly initialized from a normal distribution (with no training) thus producing six 512$\times$512 dimensional Gram matrices. Details are available in Supplementary Material\footnote{\url{https://animatedsound.com/ismir2022/metrics/supplementary/main.html}}.% Section 1.}. 
% The Gram matrix for the $n^{th}$ CNN is computed as the time-averaged correlations between the feature maps from the filters.
% Specifically, each element of the Gram matrix of the $n^{th}$ CNN is defined as 
% \begin{equation}
%     G_{pq}^n = \sum_{m=1}^M f_{pm}^n\cdot f_{qm}^n
%     = \vb{f}_p^n\cdot (\vb{f}_q^n)^T
% \end{equation}
% where, $M$ is the total number of time frames, $\vb{f}_p$ and $\vb{f}_q$ are the feature maps of dimension $1 \times M$ of the $p^{th}$ and the $q^{th}$ filters, where $p$ and $q$ range from 1 to the number of filters,$F$, i.e.~512. The resulting Gram matrix, thus, has dimensions $512\times512$ at the output of each CNN, and the Gram matrix of the $n^{th}$ CNN consists of dot products of the feature maps, written as,
% \begin{gather}
%     G = \begin{bmatrix}
% \langle \vb{f}_1,\vb{f}_1\rangle&\langle \vb{f}_1,\vb{f}_2\rangle&..&\langle \vb{f}_1,\vb{f}_F\rangle\\
% \langle \vb{f}_2,\vb{f}_1\rangle&\langle \vb{f}_2,\vb{f}_2\rangle&..&\langle \vb{f}_2,\vb{f}_F\rangle\\
% ...\\
% \langle \vb{f}_F,\vb{f}_1\rangle&\langle \vb{f}_F,\vb{f}_2\rangle&..&\langle \vb{f}_F,\vb{f}_F\rangle
% \label{eq:GM}
% \end{bmatrix}
% \end{gather}

\subsubsection{Gram matrix Cos Metric (GMcos)}
We compute the cosine distance between the Gram matrices (instead of mean square error) as,
\begin{equation}
    GMcos = 1- \frac{G_A^n\cdot {G_B^n}^T}{||G_A^n||_2||G_B^n||_2}
\end{equation}
where, $G_{A}^n$ and $G_{B}^n$ are flattened Gram matrices of audio clips A and B respectively for the $n^{th}$ CNN in an ensemble of $N$ CNNs.

\subsubsection{Accumulated-Gram Metric (AGM)}
Gram matrices for audio textures are usually sparse (see Supplementary Material (Section 1.) for more info).
% CG: Moved this to the Appendix \textcolor{Brown}{Figure \ref{fig:hist} shows the histogram of the values of the Gram matrices from the first ($k_1=2$) and the sixth ($k_6=128$) CNNs from a Frequency Modulation (FM) signal at a particular carrier frequency, modulation frequency, and modulation index and a water-filling sound at a particular fill level in a container. Although the range of values in these Gram matrices are different (x-axis), it is evident from the histograms of the Gram matrices of both these textures that they are sparse with most values of the matrix being close to zero. This suggests that non-zero values of the correlation between feature maps are sparsely located.} 
Neto et al.~\cite{neto2021improving} used a compact version of the Gram matrices to predict the class label of a given sample.
% \textcolor{Red}{They stated that since assessing every single correlation of the Gram matrix is redundant and impracticable, they aggregated the rows of the Gram matrix to achieve accumulated pairwise correlation for each feature map, which they use for classification. To design} 
Similarly, for a metric that captures the essence of the Gram matrices generated from an audio texture clip, we compute a \textit{Gram vector} corresponding to the clip by accumulating the values from its six Gram matrices, and then aggregating the rows over all the Gram matrices to get a compact one dimensional summarizing vector of length 128. 
% We choose a division factor of 4 (i.e.~512/4=128), thus a row segment is of length 128, and we average over all non-overlapping row segments of each matrix, and then over all the matrices, resulting in a gram vector of dimension $1\times128$ for any given audio texture clip. 
For more details, please refer to Supplementary Material. %\footnote{\url{https://animatedsound.com/ismir2022/metrics/supplementary/main.html}(Section 1.)}.

We define Accumulated-Gram metric (AGM) as the dot product of the difference of the two gram vectors $g_A$ and $g_B$ corresponding to the two audio clips A and B as
\begin{equation}
    AGM = (g_A-g_B)\cdot(g_A-g_B)^T
\end{equation}
\vspace{-0.2cm}
\subsection{Cochlear Param-Metric (CPM)}

McDermott and Simoncelli \cite{mcdermott2011sound} use a 3-step texture model to generate a set of statistics of an audio texture as synthesis parameters. A filterbank with a set of cochlear filters is first applied on the incoming sound to decompose the sound to many sub-bands. Then, sub-band envelopes are computed and compressed. And finally a filterbank with a set of modulation filters is applied on each compressed subband envelope. They use seven sets of time-average statistics of nonlinear functions of either the envelopes or the modulation bands as a representation of textures, which includes marginal statistics, variance, and correlations. During synthesis, a white noise signal is iteratively modified to minimize the distance between the synthesised and target sound statistics. Our experiments are based on McWalter's implementation \cite{mcwalter2018adaptive} of the \cite{mcdermott2011sound} algorithm. An overview of the method, implementation details, and statistical parameters are summarized in Supplementary Material (Section 2.). %\footnote{\url{https://github.com/rmcwalter/STSstep}}

% The implementation uses a cochlear filterbank with 36 Gammatone filters, with Hilbert transform to compute the envelopes, and then compresses the envelopes with an exponential rate of 0.3 on each element in the matrix. The modulation filterbank with 20 filters is applied on each envelope, thus generating 36$\times$20 modulation bands. A simplified overview of the method of derivation of the statistics by this method is shown in Figure \ref{fig:CPM}.

% For each sound $x$, seven sets of statistics are computed, denote them by $S^i$ where $i=[1...7]$, 
% \begin{itemize}[noitemsep,topsep=0pt]
%     % \setlength\itemsep{-0.1em}
%     \item Sub-band envelope power: average of the square sum, 36$\times$1.
%     \item Sub-band envelope marginal statistics: mean, variance, skewness, kurtosis, 36$\times$4.
%     \item Sub-band envelope variance: 36$\times$1.
%     \item Sub-band envelope correlations: the correlation between each envelope, 36$\times$36.
%     \item Modulation power: average of the square sum of each modulation bands, 36$\times$20.
%     \item Modulation variance: 36$\times$20.
%     \item Modulation correlation: the correlation between the same modulation bands for every 2-step nearest envelopes, in total (36$\times$5-6)$\times$20 = 3,480. 
% \end{itemize}
% The algorithm, thus, generates 6,432 statistical parameters.
%The code, thus, generates 6,432 statistical parameters. %The synthesis algorithm iteratively computes the distance for each statistic between the target and noise-initialized signal and uses L-BFGS algorithm to minimize the distances. 

We use the statistical parameters calculated by this algorithm as a representation for the Cochlear Param-Metric (CPM). For each sound $x$, we calculate the seven sets of statistics $S^i$ where $i=[1...7]$, and compute CPM as
% the cosine distance between each of these seven flattened statistic matrix, then add them all together as the final distance metric between different sounds, as
% This way, the metric gives equal importance to each set of statistics instead of the individual parameters in each set, thereby avoiding any bias that may be caused by the number of statistics present in each set. 
% We formally define Cochlear Param-Metric (CPM) as,
\begin{equation}
    CPM = \sum_{i=1}^{7} D(S_{A}^i, S_{B}^i)
\end{equation}
where $S_{A}^i$ is a flattened vector of the $i^{th}$ statistics set of sound clip A, and $D(x1, x2)$ calculates the cosine distance between two vectors.
\vspace{-0.2cm}
\section{Experimental Setup}
\label{sec:experiments}
\vspace{-0.2cm}
\subsection{Dataset}
\label{sec:dataset}
We use a subset of audio textures from \cite{Wyse2022Syntex} as summarized in Table \ref{tab:dataset}. There are 13 texture types, each of which has a collection of examples that are determined by a set of variable control parameters. For example, various \textit{windchimes} spunds are determined by parameters \textit{chime size} and \textit{strength}. For our experiments only one  control parameter varies at a time. For example, for the set of audio textures \textit{windchimes-strength}, we generate 11 audio files with a varying strength parameter, with the first file having the lowest strength parameter value and the 11th file having the highest strength value.  We generate 10 versions of these 11 files, where all the 10 versions have the same parameter settings but are different instances. %For example, the 10 versions of the texture with the lowest rate parameter value in the \textit{pops-rate} texture set would have the same rate, center frequency, and irregularity parameter values, but they would not be temporally identical due to the statistical variation inherent in the texture.
All the audio files being used are between 1.5 to 2 seconds long.

%The textures are divided into three groups - \textit{pitched}, \textit{rhythmic}, and \textit{others} that include non-pitched and non-rhythmic sounds. This grouping of textures was also used by \cite{antognini2019audio}, in part because they each provide different analysis and generative modeling challenges and exhibit different failure modes in models. 
The \textit{pitched} sound group consists of the frequency modulation, windchimes, chimes (without the wind sound), and feedback noise (FB noise). FB noise has the parameter `pitchedness' that changes the texture from a noisy signal to a pitched sound. Under the \textit{rhythmic} group are pops, chirps, tapping, and drum-break. The tapping sound is similar to the ``tapping 1-2'' sound in the collection of sounds in \cite{mcdermott2011sound}, and the drum break sound set is recorded and then processed with varying tempo and reverb parameters. The \textit{others} group includes wind, water filling a container, buzzing bees, and applause. Most sounds are synthesized\footnote{Dataset Appendix: \url{https://animatedsound.com/ismir2022/metrics/appendix_dataset/index.html}}, except the waterfill, Nsynth brass, and drum-break textures, which are recorded or are manipulated versions of a real recording.

\begin{table}
\small
\caption{Dataset overview}
\label{tab:dataset}
\small
    \centering
    \resizebox{!}{4cm}{\begin{tabular}[width=0.9\columnwidth]{|c|c|c|}
    \hline
    \textbf{Group} & \textbf{Texture Type} & \textbf{Parameters} \\\hline
    \hline
         &FM&modulation frequency (FM-mf)  \\
         &&carrier frequency (FM-cf) \\
         &&modulation index (FM-mi) \\\cline{2-3}
         Pitched&Windchimes&chimesize (windchimes-size)\\
         &&strength (windchimes-strength)\\\cline{2-3}
         &Chimes&chimesize (chimes-size)\\
         &&strength (chimes-strength) \\\cline{2-3}
         &FB Noise &pitchedness (fbnoise-pitchedness) \\\cline{2-3}
         &Nsynth brass&pitch (nsynth-pitch)\\\hline\midrule
         &Pops&rate (pops-rate) \\
         &&center freq (pops-cf) \\
         &&irregularity (pops-irreg) \\\cline{2-3}
         &Chirps&rate (chirps-rate) \\
         Rhythmic&& center freq (chirps-cf) \\
         && irregularity (chirps-irreg) \\\cline{2-3}
         &Tapping&rate (tapping-rate) \\
         &&relative phase (tapping-relphase) \\\cline{2-3}
         &Drum break& tempo (drum-tempo) \\
         &&reverb (drum-rev) \\\hline\midrule
         &Wind&gustiness (wind-gust)\\
         &&howliness (wind-howl)\\
         &&strength (wind-strength)\\\cline{2-3}
         Others&Waterfill&fill-level (water-fill) \\\cline{2-3}
         &Bees&center frequency (bees-cf)\\
         &&busy-body (bees-busy)\\\cline{2-3}
         &Applause&rate (applause-rate) \\
         &&no.~of clappers (applause-clappers) \\\hline
    \end{tabular}}
\end{table}
\vspace{-0.2cm}
\subsection{Subjective Evaluation}
Listening tests were conducted to quantify listeners' sensitivity to control parameter changes based on their ability to identify the order and proximity of audio samples w.r.t two selected reference samples as well as each other. This evaluation was done to provide a perceptual baseline for our objective metrics comparison. Each audio trial consisted of 7 audio samples - 2 references and 5 test samples. The 2 references were selected for each texture with the control parameters (as in the Parameters column in table \ref{tab:dataset}) set to 0 (left endpoint) and 1 (right endpoint). The 5 test samples are from the same texture selected with parameters between 0 and 1. A total of 9 such texture-trials were created, each with a different combination of control parameter settings for the 5 test audio samples. This was done to ensure that all the 9 control parameters values between 0 and 1 are included and are uniformly distributed across trials. % across all the trials for the given texture-parameter combination.
% Overall, 243 trials were created for this test.
The references, test samples, and the sequence of the individual texture-trials were randomized while conducting the test.

An interface was developed specifically for this test with the goal to intuitively convey the task details to the listeners.
% and reduce the cognitive load on individual participants from having to remember each of the 7 audio clips while answering the questions.
% Participants had to follow 6 steps before submitting a task. 
First they listened to the two references and then to the individual samples in the test. Then they were asked to create an arrangement by positioning thumbs on a slider corresponding to each test sample depending on how they perceived the order and distance of each sample w.r.t. the two references, which they could re-adjust after listening to their entire arrangement.
% The thumbs also triggered their corresponding audio sample with a double-click.
% The participants had to listen to the full arrangement in sequence and re-adjust the slider thumb positions based on their perception of the samples w.r.t each other. An instructional video was provided to guide the participants through the task. 
The listening test interface can be viewed on our webpage \footnote{\url{https://animatedsound.com/ismir2022/metrics/}}. %\footnote{The listening test interface can be found at \url{https://animatedsound.com/ismir2022/metrics/ui/question_template-with-data.html}}. 
Amazon's Mechanical Turk (AMT) was used to collect 30 responses per trial from a total of 348 unique participants. 
%Participants were allowed to attempt the task if they had a HIT (Human Intelligence Task) approval rate of 95\% and above with at least 1000 approved HITs. They were compensated at the rate of US\$0.18 for each task and a total of 7,290 responses were collected. %Participants completed on an average 21 trials with an average of 2 minutes 50 seconds per trial.
%The experimental design for these listening tests were approved by our Department's Ethics Review Committee. 
\vspace{-0.2cm}
\section{Experiments and Results}
\label{sec:results}
In this study, we experimentally address two main questions: 1) Are the objective metrics consistent for texture instances generated with the same parameters (Section \ref{sec:metricconsistency})? 2) Are the objective metrics sensitive to parameter variations (Section \ref{sec:metricsensitivity})? 
%Since the variation in the control parameter in each texture type is synthetic, we conduct a subjective listening test to determine if the control parameter variations in the different texture-parameter combinations are perceptually significant (Section \ref{sec:subjtests}), and 
We evaluate each metric by comparing them with the subjective responses. 
\vspace{-0.2cm}
\subsection{Subjective tests}
\label{sec:subjtests}
\begin{figure}
    \centering
    \includegraphics[width=0.85\columnwidth]{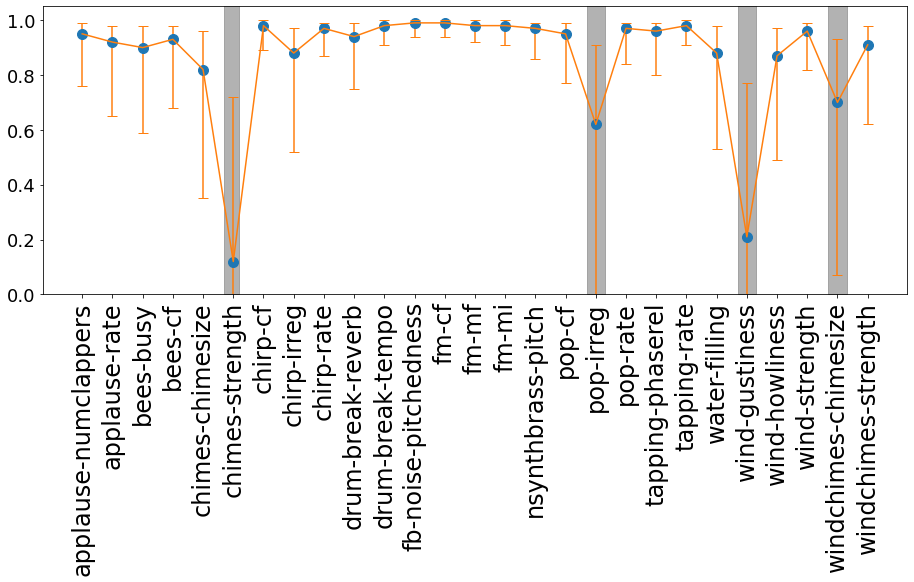}
    \vspace{-0.2cm}
    \caption{Correlation of perceptual distances with control parameters used to generate the textures. Grey bars indicate textures with no significant correlation.}
    \label{fig:human-controlparam-corr}
    % \vspace{-0.5cm}
\end{figure}

 Figure \ref{fig:human-controlparam-corr} shows the correlation coefficients for the subjective perceptual distance captured w.r.t the synthesis control parameters. Parametric variations for chimes-strength, pop-irreg, wind-gustiness and windchimes-chimesize did not result in significant correlations primarily due to wide variance in human ratings, and are excluded from further comparison with objective metrics. Based on the distance measures, we derive rank ordering for the audio samples along the parametric dimension to compare with metrics. See the Supplementary Material for summary rank-order plots obtained from the listening tests for all textures under consideration in this paper. A selection of sounds used in our listening tests can be auditioned on our webpage. %\footnote{\url{https://animatedsound.com/ismir2022/metrics/supplementary/main.html} Section 4.} 

%Based on this distance measure, we derive rank ordering for the audio samples w.r.t. to the reference audio endpoints. In our analysis, we find that using rank orders to measure differences between the samples are better suited to our comparative study requirements than using absolute perceived distance values. The absolute distances captured had large variances across all participant trials as compared to rank ordering for each texture. We thus base our objective metrics comparison using rank order of the textures. 
% Figure \ref{fig:human-test-ranks-selected} shows rank order summary plots of the listening tests for texture fm-cf, drum-break-tempo and applause-numclappers.

% \begin{figure}
%     \centering
%     \includegraphics[width=\columnwidth]{images/human-test-ranks-selected.png}
%     \caption{Shows average rank order placement captured from human listening tests for audio samples generated using the corresponding control parameters for examples in (a) pitched sounds, (b) rhythmic sounds and, (c) other textures.}
%     \label{fig:human-test-ranks-selected}
% \end{figure}
\vspace{-0.2cm}
\subsection{Metric Consistency}
\label{sec:metricconsistency}
Metric consistency means that the distance between two texture instances with the same parameter settings is  small.  We analyse consistency in terms of relative mean for three texture types - FM, pops, and water filling. The metric values are computed over one hundred comparisons between two parametrically identical textures. Relative mean is the single parameter mean with respect to the maximum average mean value of comparisons across different parameter settings for the texture as computed in Section \ref{sec:metricsensitivity}. 
% Relative standard deviation is the percentage of the standard deviation of the same parameter setting comparison values with respect to the mean of these values. This term will show the relative spread of these values, meaning how consistently is a metric able to say that the distance between textures with same parameter values is small.

The results are shown in Table \ref{tab:metricconsistency}. 
The L2 metric shows a high relative mean value for all three texture types which confirms the understanding that a spectrogram-based distance measurements does not capture the statistics that define a texture. It is too sensitive to the "irrelevant" variations between similarly parameterized textures. 

Although the relative mean of FAD is not as high as that of L2, it is higher than the Gram matrix based metrics, because the embeddings extracted from VGGish provide a holistic representation of the audio including both the overall statistics and temporal structure. The effect of the exact temporal structure of the sound on the metric is reduced but still present. The statistical metrics (CPM and GM-based) exhibit low relative mean values (good consistency), especially GMcos. The waterfill texture shows higher relative mean values across all metrics. This is at least in part because for real recordings, even for short durations, the fill-level is never constant while filling a container. 

\begin{table}
\caption{Metric Consistency in terms of relative mean in \% for three texture types - FM, pops, and water filling. Lower is better. The best two in every row are highlighted in bold.}
    \label{tab:metricconsistency}
    \centering
    \small
    \resizebox{!}{0.65cm}{
    \begin{tabular}[width=0.3\columnwidth]{|c|c|c|c|c|c|c|}
    \toprule
    \textbf{Text-Param}&\textbf{L2}&\textbf{FAD}&\textbf{CPM}&\textbf{GM}&\textbf{GMcos}&\textbf{AGM}\\\hline
        FM-cf&54.04&5.69&8.05&\textbf{0.07}&\textbf{0.02}&0.12\\\hline
        pops-rate&21.36&5.27&\textbf{3.44}&6.81&\textbf{2.46}&4.28\\\hline
        water-fill&61.09&40.60&23.09&16.19&\textbf{7.83}&\textbf{12.17}\\\hline
    \end{tabular}}
\end{table}
\vspace{-0.2cm}
\subsection{Metric Sensitivity to Parameter Variation}
\label{sec:metricsensitivity}

\begin{figure}
    \centering
    \includegraphics[width=\columnwidth]{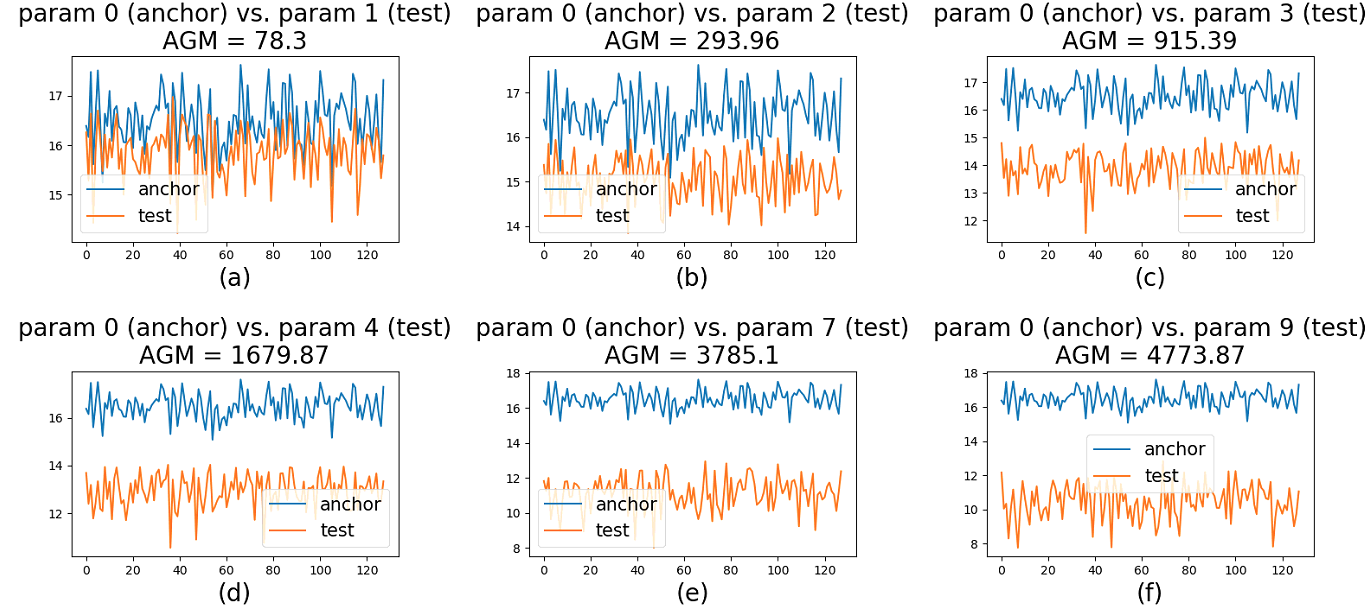}
    \vspace{-0.6cm}
    \caption{Gram vectors and AGM values of anchor for applause sounds with an increasing number of clappers from (a) to (f) compared to the anchor sound of a single clapper.}
    \label{fig:cgm_waterfill}
\end{figure}
To further investigate the sensitivity of the objective metrics to parameter variations, we fix one texture example to a low parameter value (0), called the \textit{anchor}, and compare it with nine \textit{test} clips of the same texture-type but increasing parameter values (1-9). 
%For example, for investigating the parameter sensitivity of a metric to variation in the carrier frequency (cf) parameter in FM texture-type, we compute the metrics between an anchor FM clip that has a low cf value, with nine test FM clips with gradually increasing cf values. 
We compute this over 10 versions of the same anchor and test parameter settings, and average over them for the metrics L2, CPM, GM, GMcos, and AGM. However, to calculate FAD, we compute Fréchet distance between normal  distributions of embeddings of the 10 versions, instead of between the embeddings of individual audio files (Eq.~\ref{eq:FAD}). An example of the sensitivity of the 1$\times$128 dimensional AGM Gram vector to parameter variations is shown in Figure \ref{fig:cgm_waterfill}. 
% We observe that this collapsed 1$\times$128 dimensional Gram vector is sensitive to parameter variations, as shown in Figure \ref{fig:cgm_waterfill}. 
The Gram vectors of a reference (anchor) applause audio texture clip that has the number of clappers parameter set to 1, is compared with six test audio clips of applause textures with increasing number of clappers. The divergence between the Gram vectors of the anchor and test sounds grows as the number of clappers is increased.

\begin{table}
\centering
\caption{Pearson's Correlation between the avg human rank orders and the distance from the anchor computed by the objective metrics. The best two values in every row are in bold. Correlation values >=0.5 are green, <0.5  orange.}
\label{tab:comparisonRank}
    \centering
    \small
    \resizebox{!}{4cm}{\begin{tabular}[width=0.9\columnwidth]{|c|c|c|c|c|c|c|}
% \label{tab:comparisonRank}
% \begin{tabular}{|c|c|c|c|c|c|c|} 
\hline
\textbf{Texture-Param} & \textbf{L2}                       & \textbf{FAD}                      & \textbf{CPM}                     & \textbf{GM}                               & \textbf{GMcos}                    & \textbf{AGM}                       \\ 
\hline
\multicolumn{7}{|c|}{\textbf{Pitched}}                                                                                                                                                                                                                 \\ 
\hline
FM-mf                  & {\cellcolor{green}}\textbf{0.99}  & {\cellcolor{green}}0.94           & {\cellcolor{green}}\textbf{0.99} & {\cellcolor{green}}\textbf{1.00}          & {\cellcolor{green}}\textbf{0.99}  & {\cellcolor{green}}0.64            \\ 
\hline
FM-cf                  & {\cellcolor{green}}\textbf{0.99}  & {\cellcolor{green}}0.71           & {\cellcolor{green}}\textbf{0.99} & {\cellcolor{green}}\textbf{1.00}          & {\cellcolor{green}}\textbf{0.99}  & {\cellcolor{green}}0.97            \\ 
\hline
FM-mi                  & {\cellcolor{green}}\textbf{0.99}  & {\cellcolor{green}}0.75           & {\cellcolor{green}}0.94          & {\cellcolor{green}}\textbf{0.99}          & {\cellcolor{green}}\textbf{1.00}  & {\cellcolor{green}}\textbf{0.99}   \\ 
\hline
windchimes-strength    & {\cellcolor{green}}\textbf{0.97}  & {\cellcolor{green}}\textbf{0.97}  & {\cellcolor{green}}0.82          & {\cellcolor{green}}\textbf{0.95}          & {\cellcolor{green}}\textbf{0.95}  & {\cellcolor{green}}0.62            \\ 
\hline
chimes-size            & {\cellcolor[rgb]{1,0.647,0}}0.28  & {\cellcolor{green}}\textbf{0.88}  & {\cellcolor{green}}\textbf{0.92} & {\cellcolor[rgb]{1,0.647,0}}0.20          & {\cellcolor[rgb]{1,0.647,0}}0.20  & {\cellcolor[rgb]{1,0.647,0}}0.07   \\ 
\hline
fbnoise-pitchedness    & {\cellcolor{green}}\textbf{1.00}  & {\cellcolor{green}}0.75           & {\cellcolor{green}}0.98          & {\cellcolor{green}}\textbf{1.00}          & {\cellcolor{green}}\textbf{0.99}  & {\cellcolor{green}}\textbf{1.00}   \\ 
\hline
nsynth-pitch           & {\cellcolor[rgb]{1,0.647,0}}0.09  & {\cellcolor[rgb]{1,0.647,0}}-0.45 & {\cellcolor{green}}\textbf{0.73} & {\cellcolor[rgb]{1,0.647,0}}\textbf{0.23} & {\cellcolor[rgb]{1,0.647,0}}0.09  & {\cellcolor[rgb]{1,0.647,0}}-0.50  \\ 
\hline
\multicolumn{7}{|c|}{\textbf{Rhythmic}}                                                                                                                                                                                                                \\ 
\hline
pops-rate              & {\cellcolor{green}}\textbf{0.98}  & {\cellcolor{green}}0.85           & {\cellcolor{green}}0.80          & {\cellcolor{green}}0.79                   & {\cellcolor{green}}0.89           & {\cellcolor{green}}\textbf{0.94}   \\ 
\hline
pops-cf                & {\cellcolor{green}}\textbf{0.98}  & {\cellcolor{green}}0.96           & {\cellcolor{green}}\textbf{0.99} & {\cellcolor{green}}\textbf{0.99}          & {\cellcolor{green}}\textbf{0.99}  & {\cellcolor{green}}\textbf{0.98}   \\ 
\hline
chirps-rate            & {\cellcolor{green}}\textbf{0.97}  & {\cellcolor{green}}\textbf{0.94}  & {\cellcolor{green}}\textbf{0.94} & {\cellcolor{green}}0.84                   & {\cellcolor{green}}0.88           & {\cellcolor{green}}0.89            \\ 
\hline
chirps-cf              & {\cellcolor{green}}\textbf{0.99}  & {\cellcolor{green}}\textbf{0.98}  & {\cellcolor{green}}\textbf{0.98} & {\cellcolor{green}}\textbf{0.98}          & {\cellcolor{green}}\textbf{0.98}  & {\cellcolor{green}}0.97            \\ 
\hline
chirps-irreg           & {\cellcolor{green}}0.81           & {\cellcolor{green}}\textbf{0.95}  & {\cellcolor{green}}\textbf{0.94} & {\cellcolor{green}}0.90                   & {\cellcolor{green}}0.92           & {\cellcolor{green}}0.90            \\ 
\hline
tapping-rate           & {\cellcolor{green}}\textbf{1.00}  & {\cellcolor{green}}0.90           & {\cellcolor{green}}0.91          & {\cellcolor{green}}0.64                   & {\cellcolor{green}}\textbf{0.94}  & {\cellcolor{green}}0.76            \\ 
\hline
tapping-relphase       & {\cellcolor{green}}0.88           & {\cellcolor{green}}\textbf{0.96}  & {\cellcolor{green}}\textbf{0.98} & {\cellcolor{green}}0.94                   & {\cellcolor{green}}0.95           & {\cellcolor{green}}0.76            \\ 
\hline
drum-tempo             & {\cellcolor[rgb]{1,0.647,0}}0.08  & {\cellcolor{green}}\textbf{0.82}  & {\cellcolor{green}}\textbf{0.97} & {\cellcolor{green}}\textbf{0.97}          & {\cellcolor{green}}0.63           & {\cellcolor{green}}0.81            \\ 
\hline
drum-rev               & {\cellcolor{green}}\textbf{0.93}  & {\cellcolor{green}}0.85           & {\cellcolor{green}}0.81          & {\cellcolor{green}}0.90                   & {\cellcolor{green}}0.81           & {\cellcolor{green}}\textbf{0.94}   \\ 
\hline
\multicolumn{7}{|c|}{\textbf{Others}}                                                                                                                                                                                                                  \\ 
\hline
wind-howl              & {\cellcolor{green}}\textbf{0.92}  & {\cellcolor{green}}\textbf{0.92}  & {\cellcolor{green}}\textbf{0.92} & {\cellcolor{green}}0.80                   & {\cellcolor{green}}\textbf{0.92}  & {\cellcolor{green}}0.86            \\ 
\hline
wind-strength          & {\cellcolor{green}}\textbf{0.96}  & {\cellcolor{green}}0.86           & {\cellcolor{green}}\textbf{0.88} & {\cellcolor{green}}0.87                   & {\cellcolor{green}}\textbf{0.88}  & {\cellcolor{green}}0.56            \\ 
\hline
water-fill             & {\cellcolor[rgb]{1,0.647,0}}0.39  & {\cellcolor[rgb]{1,0.647,0}}0.32  & {\cellcolor[rgb]{1,0.647,0}}0.41 & {\cellcolor{green}}\textbf{0.75}          & {\cellcolor{green}}\textbf{0.73}  & {\cellcolor[rgb]{1,0.647,0}}-0.27  \\ 
\hline
bees-cf                & {\cellcolor{green}}\textbf{0.97}  & {\cellcolor{green}}0.90           & {\cellcolor{green}}\textbf{0.98} & {\cellcolor{green}}\textbf{0.97}          & {\cellcolor{green}}\textbf{0.98}  & {\cellcolor{green}}0.80            \\ 
\hline
bees-busy              & {\cellcolor[rgb]{1,0.647,0}}-0.21 & {\cellcolor{green}}\textbf{0.89}  & {\cellcolor{green}}\textbf{0.89} & {\cellcolor[rgb]{1,0.647,0}}-0.34         & {\cellcolor[rgb]{1,0.647,0}}-0.43 & {\cellcolor{green}}\textbf{0.92}   \\ 
\hline
applause-rate          & {\cellcolor{green}}0.66           & {\cellcolor{green}}0.83           & {\cellcolor{green}}\textbf{0.86} & {\cellcolor{green}}0.66                   & {\cellcolor{green}}0.78           & {\cellcolor{green}}\textbf{0.90}   \\ 
\hline
applause-clappers      & {\cellcolor{green}}\textbf{0.97}  & {\cellcolor{green}}0.94           & {\cellcolor{green}}0.93          & {\cellcolor{green}}\textbf{0.99}          & {\cellcolor{green}}\textbf{0.99}  & {\cellcolor{green}}\textbf{0.99}   \\
\hline
\end{tabular}}
\end{table}

\begin{figure}[h]
\centering
    \includegraphics[width=0.65\columnwidth]{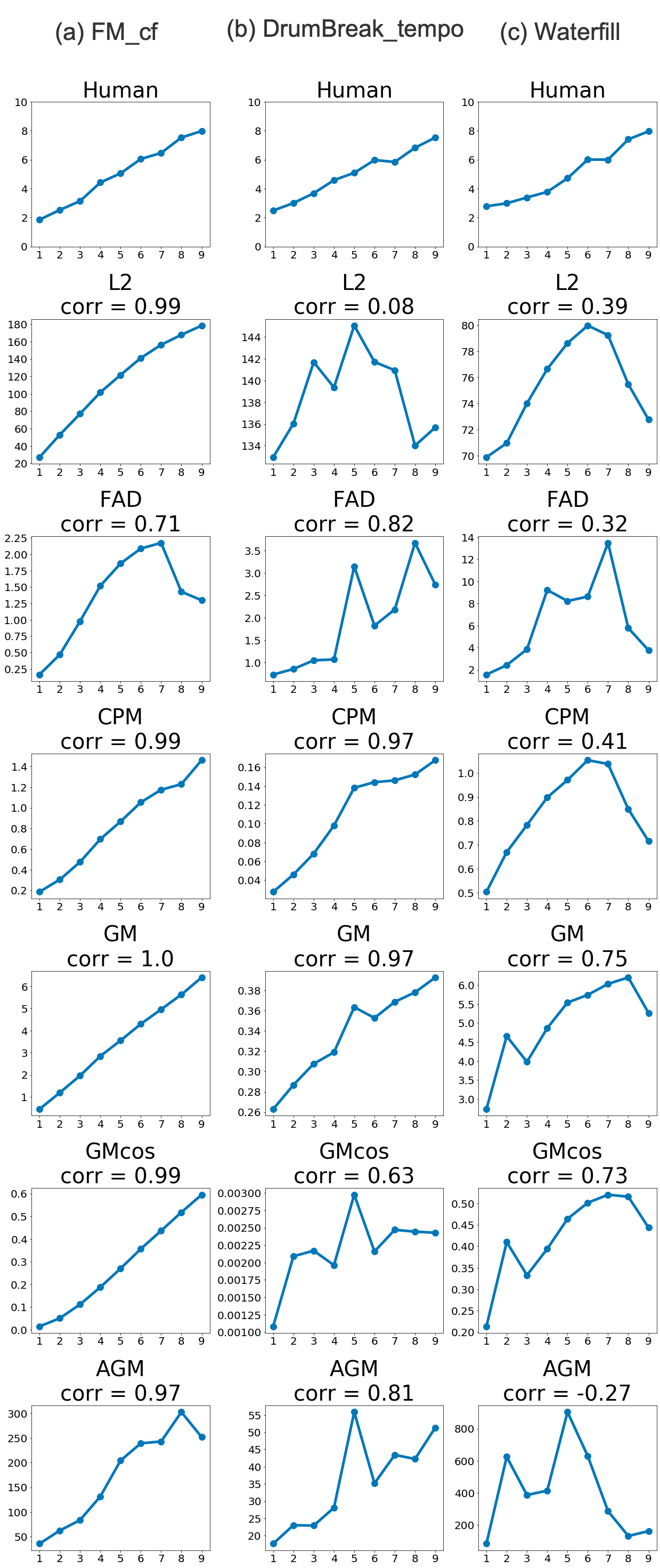}
    \vspace{-0.3cm}
    \caption{Trends of human responses and objective metrics to parameter variations for three texture-param combinations, (a) FM with carrier frequency variable parameter, (b) Drum-break with tempo a variable parameter, and (c) Waterfill with fill-level parameter, along with the Pearson's correlation between the objective metric values and the subjective responses.}
    \label{fig:maincomparison}
\end{figure}

Figure \ref{fig:maincomparison} shows the plots of all the metric values for three textures compared with human responses in terms of average rank order. The Pearson's correlation between the objective metrics and the subjective responses are presented for all textures in Table \ref{tab:comparisonRank}, and their corresponding plots are available in the Supplementary Material.%\footnote{\url{https://animatedsound.com/ismir2022/metrics/supplementary/main.html} Section 3.}.
\vspace{-0.2cm}
\subsubsection{Metric Performances}
Again, L2 metric is a poor performer. This is particularly clear for drumbreak-tempo textures as can be observed in the L2 column in Figure \ref{fig:maincomparison}(b). FAD performs better than the Gram matrix based measures for chimes-size, but shows worse performance for FM-cf, FM-mi, FBnoise-pitchedness, and waterfill (Table \ref{tab:comparisonRank}).
For most of the rhythmic textures, FAD performs well, except pop-rate, drum-tempo, and drum-rev. As FAD also preserves some information about exact spectro-temporal structure, the result is variable sensitivity to parameters (Figure \ref{fig:maincomparison}, FAD row).
% FAD metric performs well except for FM-cf, FM-mi, fbnoise-pitchedness, nsynth-pitch, and waterfill. However, some of the other texture-param combinations also show  non-monotonically increasing, or noisy FAD trend, even though the correlation values are relatively high, such as drumbreak-tempo, applause-rate, and wind-strength.

% \subsubsection{Performance of CPM}
The statistics designed by \cite{mcdermott2011sound} are known to synthesize good quality sounds for natural textures such as bees and wind which is clearly reflected in our results. However, these statistics are also known for low realism scores for synthesized pitched sounds such as windchimes, and rhythmic sounds such as drum break. Compared to the Gram matrix based metrics, CPM performs well except for windchimes-strength, pop-rate, drum-rev, and water-fill. 

% \subsubsection{Gram matrix based Metrics}
% Bees-busy is a case where AGM outperforms all other metrics, but surprisingly GM and GMcos too. As the busy-body parameter increases, not only do the buzzing vibrations become faster but their amplitude modulation increases, almost sounding like the \textit{rate} of buzzing has increased.\textcolor{red}{[Why is this notable?]} Some other cases where AGM performs better than GM are applause-rate, pop-rate, tapping-rate, and chirp-rate. When an event rate varies, the features captured by the Gram matrix may reflect some individual events, especially for shorter kernel sizes. In such cases, AGM captures the summary of those statistics instead of individual events, thereby showing a higher parameter sensitivity than metrics based on the entire matrix. A similar trend is also observed for GMcos where it outperforms GM in the cases of applause-rate, pop-rate, and chirp-rate, showing that GMcos is also influenced more by the overall temporal statistics than individual events.

When an event rate varies, the features captured by the Gram matrix may reflect some individual events, especially for shorter kernel sizes. In such cases, AGM captures the summary of those statistics instead of individual events, thereby showing a higher parameter sensitivity than metrics based on the entire matrix. This trend can be seen in applause-rate, pop-rate, tapping-rate, chirp-rate, as well as bees-busy (increase in busy body parameter sounds like increase in the rate of buzzing vibrations), where AGM outperforms GM. %A similar trend is also observed for GMcos where it outperforms GM in the cases of applause-rate, pop-rate, and chirp-rate, showing that GMcos is also influenced more by the overall temporal statistics than individual events.

The cases where GM outperforms AGM are FM-mf, bees-cf, windchimes-strength, waterfill, and wind-strength, where the parameter variation occurs in the frequency domain, i.e.~the center frequency in bees-cf, wind and windchimes-strength, modulation frequency in FM-mf, and resonant frequencies in water-fill. AGM summarization loses information about the statistics in the frequency domain, and this suggests a future study where methods such as eigenvalue decomposition might be better for extracting key information from the Gram matrix.

\subsubsection{Some curious cases}
Water-fill shows an interesting trend in most of the metrics, where there's a gradual increase, and then a decrease in the metric values (Figure \ref{fig:maincomparison}(c)), possibly because the resonances of the container wrt changing water and air column heights cause
% when the water just starts pouring may have some similarity to the resonances of an almost full container as the height of the water column and air column almost swap between these two situations. So possibly, 
the audio textures of an almost full container to have some statistical similarities with that of an almost empty container. That is, the high-level fill-level parameter has a non-linear affect on the resonant frequencies of the system. For a sound like water-fill where multiple perceptual dimensions are varying, GM and GMcos show higher correlation with humans than others.

%Drumbreak-tempo is another interesting case where except for CPM and GM, all other metrics show a noisy trend. The statistical features captured by these metrics have a large enough time context to measure the repetitions in drum loops. The AGM metric loses some temporal context information because of the summarization of Gram matrix.

Nsynth-pitch is a curious example where none of the metrics perform well except CPM. The sustained portion of musical tones are atypical as textures because of the lack of temporal variation. Each spectrogram frame is almost exactly the same. Also, as the pitch of the brass instrument increases, the harmonics also change. The Gram matrix summarizes the temporal statistics through the 1D audio representations, but not the frequency statistics, thereby showing a noisy performance.%When there is a single tone or a sound with a center frequency and no harmonics, Gram matrix based measures perform well. However, sounds that have multiple frequency components such as a musical instrument, Gram matrix based methods fail.% as they don't calculate statistics across frequencies. 
\vspace{-0.3cm}
\section{Discussion}
We explore a variety of metrics for use with audio textures for their sensitivity to controlled parametric variation. However, in a realistic situation, there may effectively be multiple simultaneous dynamic parameters. One such example in our dataset was the waterfill texture-type, where the fill parameter maps to both rising and falling resonant frequencies. We saw that most of the metrics were responding to a combination of fill parameter and resonances. Further investigation is required to understand how metrics behave in such complex realistic situations. 

The systematic parameter variation we used creates textures perceptually close to each other. Further study is required to understand the behavior of these metrics for cross texture-type comparisons, for example, bees compared to water. Such a study would help in mapping audio textures to a perceptual space, the way musical instrument space has been mapped \cite{grey1977multidimensional,esling2018bridging}.

Our perceptual study involved placing textures within a 1D space between two reference textures, but there is an inherent lack of an absolute perceptual frame of reference for the meaning of control parameter variation. A systematic study is needed to understand perceptual ``just noticeable differences'' in parameter variations for complex sounds. Different parametric variations could then be compared on a unified scale. 

% \textcolor{Brown}{
% As shown by Turian et al.~\cite{turian2020im}, many commonly used distance metrics are unable to sense the pitch direction, that is whether the pitch of a sinusoid is lower or higher than that of another sinusoid. The metrics studied in this work also do not provide any indication regarding the direction of a parameter change, which need to be studied in the future.}
% old figure 4
\vspace{-0.5cm}
\section{Conclusions}
\label{sec:conclusion}
We present a comprehensive study on the parameter change sensing property of various existing audio evaluation metrics as well as three potential audio statistical and deep-feature based metrics.%that are shown to be sensitive to parameter variations in many audio texture examples
Our code-base for the proposed metrics can be found on our website. CPM and FAD emerge as the best metrics, while GM based metrics show promising results. This shows the potential of these deep-features for the purpose of evaluation of audio textures. This study is a fruitful first step towards understanding audio texture, metric design for audio textures, building better synthesis models of this rich and complex class of sounds, and generally toward mapping the space of audio textures.
%\footnote{Our code-base for the proposed metrics can be found here: \url{https://animatedsound.com/ismir2022/metrics/code/source.zip}}
\bibliography{ISMIRtemplate}

% For non bibtex users:
%\begin{thebibliography}{citations}
% \bibitem{Author:17}
% E.~Author and B.~Authour, ``The title of the conference paper,'' in {\em Proc.
% of the Int. Society for Music Information Retrieval Conf.}, (Suzhou, China),
% pp.~111--117, 2017.
%
% \bibitem{Someone:10}
% A.~Someone, B.~Someone, and C.~Someone, ``The title of the journal paper,''
%  {\em Journal of New Music Research}, vol.~A, pp.~111--222, September 2010.
%
% \bibitem{Person:20}
% O.~Person, {\em Title of the Book}.
% \newblock Montr\'{e}al, Canada: McGill-Queen's University Press, 2021.
%
% \bibitem{Person:09}
% F.~Person and S.~Person, ``Title of a chapter this book,'' in {\em A Book
% Containing Delightful Chapters} (A.~G. Editor, ed.), pp.~58--102, Tokyo,
% Japan: The Publisher, 2009.
%
%
%\end{thebibliography}

\newpage
\appendix
\onecolumn
%%\input{SoundDataset}
%%%%%%%%%%%%%%%%%%%%%%%%%%%%%%%%%%%%%%%%%%%%%%%%%%%%%%%%%%%%%%%%%%%%%%%%%%%%%%%
%%%%%%%%%%%%%%%%%%%%%%%%%%%%%%%%%%%%%%%%%%%%%%%%%%%%%%%%%%%%%%%%%%%%%%%%%%%%%%%

\end{document}